\documentclass[12pt,epsf]{article}
\input epsf.sty
\def\bea{\begin{eqnarray}} \def\eea{\end{eqnarray}} \def\ba{\begin{array}}
\def\be{\begin{equation}} \def\ee{\end{equation}} 
\def\bi{\begin{itemize}} \def\ei{\end{itemize}} 
\def\ea{\end{array}} \def\ben{\begin{enumerate}} \def\een{\end{enumerate}}

\newcommand{\eqn}[1]{(\ref{#1})}

\newcommand{\hepth}[1]{{\tt hep-th/{#1}}}

\def\m{\mu}

\def\n{\nu}

\def\br{\nonumber\\}

\begin{document}
{}~
\hfill\vbox{\hbox{hep-th/1804.nnnn} \hbox{\today}}\break

\vskip 3.5cm
\centerline{\large \bf  RG flows and cascades of 
$Lif_4^{(2)}\times S^1\times S^5$ vacua}

\vskip 1cm

\centerline{\bf  Harvendra Singh}
\vspace*{.2cm}

\centerline{ \it  Theory Division, Saha Institute of Nuclear Physics} 
\centerline{ \it  1/AF, Bidhannagar, Kolkata 700064, India}
\vspace*{.25cm}
\centerline{ \it  and}
\vspace*{.25cm}
\centerline{ \it  Homi Bhabha National Institute, Anushakti Nagar, Mumbai 400094, India}
\vspace*{.25cm}

\vskip.5cm

\vskip1cm

\centerline{\bf Abstract} \bigskip
The   (F1,D2,D8) brane configuration produces  
$Lif_4^{(2)}\times {S}^1\times S^5$ 
 Lifshitz vacua supported by  `massive' $B$-field. 
We present  exact deformations of this system under which new 
massless $B$-modes of strings also get excited. Due to these massless 
modes the deformed solutions flow  to  conformally
 $Lif_4^{(3)}\times S^1\times S^5$ vacua in the IR.
The latter types are  
supersymmetric solutions of ordinary
 type IIA theory. The massive and massless $B_{\mu\nu}$ modes 
segregate out in the IR.
 We confirm that the massive $B$ mode and cosmological constant 
 indeed decouple from the theory  rendering the IR field dynamics
controlled by  massless fields only. A
similar effect is  observed on the UV side of the flow
where a relativistic  regime reappears. 
We also present  `cascading' Lifshitz vacua in which dynamical exponent 
has integral jumps along  the  flow; 
$Lif_4^{(3)}\to Lif_4^{(2)}\to Lif_4^{(1)}$. The  critical $Lif_4^{(2)}$ theory
separates  `deconfining'  $Lif_4^{(3)}$ IR theory  from  
the confining Yang-Mills  phase in UV.

\vfill 
\eject

\baselineskip=16.2pt

\section{Introduction}
The  AdS-CFT holography  \cite{Maldacena:1997re,Gubser:1998bc, Witten:1998qj}
has produced a nonrelativistic version of itself where
 strongly coupled quantum theories at critical points
 have been the focus of several studies  \cite{son}-\cite{taylor}.  
These systems may involve   strongly coupled fermions 
 at finite density or some gas of ultra-cold atoms 
\cite{son, bala}. 
In some studies involving 
nonrelativistic Schr\"odinger spacetimes the 4-dimensional
 bulk spacetime geometry generically requires supporting
 Higgs like  fields such as massive vector field 
 \cite{herzogrev,denefrev,son}. We particularly discuss here  
separate examples of 10-dimensional Lifshitz spacetimes
 where  Higgs like matter field is  2-rank 
antisymmetric tensor field which couples to 
excited massive modes of the strings. The two phenomena 
indeed are parallel from  higher dimensional perspective, 
because a Kaluza-Klein compactification of 2-rank tensor field
on $S^1$  gives rise to vector field in one lower dimensions.

In this work our aim is to study RG flows of the 
known $Lif_4^{(2)}\times S^1\times S^5$ solution, 
a Lifshitz vacua with dynamical exponent of time as $a=2$, 
in the `massive'  type IIA theory  \cite{roma}.
The massive type IIA string theory (mIIA) is  unique
10-dimensional maximal supergravity where the string 
$B$ field is explicitly massive and the theory also
includes a positive cosmological constant. Because of
this mIIA theory provides an unique setup to study Lifshitz  
and Schr\"odinger  like nonrelativistic
 solutions, as these  require massive gauge fields. 
The Lifshitz bulk spacetimes
 provide a holographic dual description of nonrelativistic 
quantum theories living on the
spacetime boundary \cite{kachru}. 
The recently found $Lif_4^{(2)}\times S^1\times S^5$
solution is a background 
generated by a bound state of $(F1,D2,D8)$ branes \cite{hs2017}. 
The massive T-dual Lifshitz solutions have been shown to exist 
in type IIB  theory with  axionic flux  switched 
on \cite{Balasubramanian:2010uk}.  
It is very important to study RG flows of these vacua because 
a  compactification  on $S^1\times S^5$ 
immediately provides us the prototype  
$Lif_4^{(2)}$ background in four dimensions which is   
 holographically dual to 3-dimensional Lifshitz theory. 
The RG flow
will tell us how the physical changes occur in the theory at fundamental 
level at different energy scales.  
We shall  study  deformations of 
$Lif_4^{(2)}\times S^1\times  S^5$ background,
which form exact solutions of mIIA and explicitly involve massive 
$B$-field excitations. 
The excitations  induce running of string coupling as well. It is observed
 that the resulting RG flow to deep IR (and towards UV) can  be described by
ordinary type IIA string theory (oIIA) alone. The main reason that 
this to  happen is that the 
contributions of the mass  and the cosmological constant terms
 disappear from the field
 dynamics far away from the critical point. 

The paper is organised as following. In section-2 we first review the fixed
point Lifshitz solution $Lif^{(2)}_4\times S^1\times S^5$ in massive type 
IIA theory. In section-3 we write down its deformed version
 where strings become excited. We discuss the RG flow
 where the string massive mode decouples in the IR and  conformally
$Lif^{(3)}_4\times S^1\times S^5$ vacua of ordinary type IIA appear as spacetime. 
The section-4
containes  UV deformations of the $Lif^{(2)}_4\times S^1\times S^5$ solutions.
In the UV regime too the massive string field and cosmological term decouple.
In section-5, we  present a cascading type solution. Here we find that 
the difference $(a-\theta)$ remains constant along the RG flow 
through out the cascade. 
The results are summarised in section-6.
  
\section{$Lif^{(2)}_4\times S^1\times S^5$ vacua}
 
 The Romans type IIA theory is the only known maximal 
supergravity in ten dimensions which allows massive string $B_{\mu\nu}$  
field. The theory is described  by the following bosonic action
\bea
S={1\over G_N}\int  \bigg[e^{-2\phi}\left(  \ast R + 4
(d\phi)^2 
-{1\over 2} (H_{(3)})^2  \right)
-{1\over 2} (G_{(2)})^2  
-{1\over 2} (G_{(4)})^2  
-\ast{m^2\over 2}  \bigg]
\eea
where  topological terms have been dropped because
these would be  vanishing 
for the Lifshitz backgrounds we are studying in this paper, see 
for details in \cite{roma, hs2001}.\footnote{
We are adopting a convention: $
\int (H_{(p)})^2=\int H_{(p)}\wedge\ast H_{(p)}
={1\over p!}\int d^{10}x \sqrt{-g}
H_{\mu_1\cdots\mu_p}H^{\mu_1\cdots\mu_p}$ and for  scalar quantities 
like curvature scalar: 
$\int \ast R=\int d^{10}x \sqrt{-g} R$  .}  The field strengths 
are defined as
\bea
H_{(3)}=dB_{(2)},~~~G_{(2)}=dC_{(1)} +m B_{(2)}, ~~~
G_{(4)}=dC_{(3)} +B_{(2)}\wedge dC_{(1)} + {m\over2} 
B_{(2)}\wedge B_{(2)}
\eea
where $m$ is the mass parameter and $m^2$ a positive cosmological constant of
the 10-dimensional theory. The cosmological constant 
 generates   a nontrivial  potential term  for the dilaton 
field. Other than the well known  
Freund-Rubin $AdS_4\times S^6$  vacua  \cite{roma}, some 
  supersymmetric solutions of the  theory 
include D8-branes \cite{Pol1, berg, wittd0d8,park00,ohta01,hull}, the
$K3$ compactifications \cite{haack}, the $(D6,D8)$  and  $(D4,D6,D8)$ 
bound states \cite{hs2001, hs1}.
Under the `massive' T-duality \cite{berg} the D8-brane (domain walls)  
 can be mapped  to  D7-brane of type IIB  theory.  
The string $B$-field  is explicitly massive  
and it plays  important role in obtaining
nonrelativistic $a=3$ Schr\"odinger solutions \cite{Singh:2009tq}. 
The massive type IIA theory however  never admits a  Minkowski 
solution. 
   
An observed   feature in four-dimensional AdS gravity theories
has been that in order to obtain 
Schr\"odinger or Lifshitz type non-relativistic 
solutions one needs to include a
massive (Proca like) gauge fields in the theory \cite{son,bala}. 
(Although massless gauge fields  can give rise to  nonrelativistic vacua 
however,  in  simple cases of  D$p$-branes
compactified along a worldvolume  lightcone coordinate \cite{hs2010}, 
these solutions straight forwardly give rise to  conformal (or hyperscaling) 
Lifshitz and Schr\"odinger vacua \cite{hs2012}.)

\subsection{F1-D2-D8 system:  Lifshitz vacua and massive strings}

The  $AdS_4\times S^6$ maximally symmetric
 Freund-Rubin vacua in Romans' theory \cite{roma} 
is constituted by D2 and D8-branes. The D8-brane couples to a 10-form 
field strength $F_{(10)}=\star m$, where $m$ is the cosmological constant
in massive type IIA supergravity. 
It has  recently been  shown  that one can even construct  Lifshitz   vacua 
which are  constituted by  `massive' strings, D2 and D8 branes \cite{hs2017}.
The string field in the Lifshitz solutions becomes massive after gobbling up  
the D0-branes. Thus massive $B$-field carries additional degrees of freedom
as compacted to the massless one.   
The $Lif_4^{(2)}\times S^1\times S^5$ Lifshitz
 solutions are given by 
(in string  metric and $\alpha'=1$) 
\bea\label{sol2a}
&&ds^2= L^2\left(- {dt^2\over  z^4} +{dx_1^2+dx_2^2\over z^2}+{dz^2\over
z^2}  +{dy^2\over  q^2} + d\Omega_5^2 \right) ,\br
&&e^\phi=g_a, ~~~~~C_{(3)}= -{L^{3} \over g_a}
{1\over  z^4}dt\wedge dx_1\wedge dx_2, \br &&
B_{(2)}=  { L^2\over q z^2}dt\wedge dy  \ ,
\eea 
with $L$ being related to $m$ as $L={2\over g_a  m l_s}$, 
and $m$ being the mass parameter in the mIIA action 
and  equations of motion.
 The constant $q$ is a free (length)  parameter 
and $g_a$ we take to be weak string coupling $(g_a<1)$
in this  massive type IIA vacuum. 
Note $L$, which is dimensionless,
determines overall radius of curvature of the 10-dimensional spacetime.
While $m$  a parameter of the lagrangian theory is related to 
 $L$. Therefore Romans' theory with 
  $m \ll{ 2\over g_a l_s}$ would be preferred 
so that we can have $L\gg 1$ in  the solutions \eqn{sol2a},  else
  these classical vacua cannot be trusted.\footnote{Note, from  the D8
brane/domain wall idea discussed in \cite{berg}, one typically
expects $m \approx {g_a N_{D8} \over l_s}$, a value which is definitely
well within ${ 2\over g_a l_s}$ for finite number, $N_{D8}$, of $D8$ branes
in the background.}

The Lifshitz configuration \eqn{sol2a} describes  parallel stack 
of D2-branes with brane directions stretched along $(x_1,~x_2)$
 and  `massive' fundamental strings that are aligned
along  $y$ direction. The D8-branes  would wrap around $S^5$ 
 and remaining part of the worldvolume 
is stretched along the  patch $(x_1,~x_2,~y)$.

\section{ $Lif_4^{(2)}\times S^1\times S^5$ vacua 
 with IR deformations}

The $a=2$ Lifshitz vacua allows following type of  deformations 
\bea\label{sol2a8}
&&ds^2= L^2\left(- {dt^2\over  z^4h} +{dx_1^2+dx_2^2\over z^2}+{dz^2\over
z^2}  +{dy^2\over  q^2h} + d\Omega_5^2 \right) ,\br
&&e^\phi=g_a h^{-1/2},
 ~~~~~C_{(3)}= -{L^{3} \over g_a}
{1\over  z^4}dt\wedge dx_1\wedge dx_2, \br &&
B_{(2)}=  { L^2\over q  z^2}h^{-1}dt\wedge dy  \ ,
\eea 
where the function $$h= 1+{z^2\over z_{0}^2}. $$ 
The excitations  involve 
$(t,y)$ metric components, leaving the  $x_1,x_2$ 
plane of D2-branes 
unaffected. These  excitations also induce a
running  dilaton field. 
The string $B$-field along $y$ direction 
also gets coupled to the  excitations. Since
$h\sim 1$ near  $z= 0$, 
these excitations are normalizable modes (usually $z_0$ corresponds to 
switching on relevant operators in the Lifshitz theory). It is clear that the 
solution \eqn{sol2a8} flows to a weakly coupled  
 fixed point solution \eqn{sol2a} in the UV. The string coupling
always stays weak so long as $g_a<1$ even with the deformations.

The vacua \eqn{sol2a8}  form exact solutions of the m-IIA theory. 
This  enables us to also study the  IR region of 
 $a=2$  Lifshitz theory. In the
deep IR region, $z\gg z_0$, where $h\approx {z^2\over z_0^2}$,
the  Lifshitz vacua is  driven to a  
weakly coupled  regime.  For $z\gg z_0$,  the  IR geometry becomes a  conformally
  $Lif_4^{(3)}\times S^1\times S^5$ solution: 
\bea\label{sol2a9}
&&ds^2_{IR}\sim L^2\left(- {z_0^2dt^2\over  z^6} +{dx_1^2+dx_2^2\over z^2}+{dz^2\over
z^2}  +{z_0^2 dy^2\over  z^2 q^2 } + d\Omega_5^2 \right) ,\br
&&e^\phi\sim g_a {z_0\over z},  ~~~~~C_{(3)}= -{ L^3 \over g_a}
{1\over  z^4}dt\wedge dx_1\wedge dx_2, \br &&
B_{(2)}\sim  { L^2 z_0^2\over q z^4}dt\wedge dy  \ .
\eea 
In the IR regime 
the $y$ circle essentially becomes smaller. (Once it becomes substringy,
in that situation we should opt to go to
type IIB dual theory that will provide a better holographic description.)
Otherwise also, if  $y$ coordinate is a flat noncompact direction,  
 then the IR geometry would resembles $Lif_4^{(3)}\times R^1\times S^5$ 
spacetime describing a 4-dimensional $a=3$ Lifshitz theory 
on its boundary at $z=0$. Along this IR flow
the massive  modes of  bulk fields  completely decouple in deep IR region
leaving behind a massless field description, at very weak string coupling.
Due to weak string coupling fluctuations are suppressed near $z\sim\infty$. 
(This may indicate the presence of a horizon.)

The constant $z_0$ in \eqn{sol2a9} has no particular 
importance as it can be removed by  redefining 
$g_a, t, y$ suitably. For example, with the definitions: $g_a'=g_a z_0,~
 t'=z_0 t,~ x_3={z_0\over q} y$, we can get 
\bea\label{sol2a9n}
&&ds^2= L^2\left(- {(dt')^2\over  z^6} +{dx_1^2+dx_2^2\over z^2}
 +{dz^2\over z^2}+{dx_3^2\over z^2}
   + d\Omega_5^2 \right) ,\br
&&e^\phi={g_a' \over z},  ~~~~~C_{(3)}= -{ L^3 \over g_a'}
{1\over  z^4}dt'\wedge dx_1\wedge dx_2, \br &&
B_{(2)}=  { L^2 \over z^4} dt'\wedge dx_3  \ ,
\eea 
It is required to note  that conformally
  $Lif_4^{(3)}\times S^1\times S^5$ background 
 \eqn{sol2a9n} are  described by 
the  ordinary type IIA string theory, 
of which this is an exact  solution. (These same solutions
appear in the eq.(20) of the work \cite{hs2010}.
The $SO(3)$  symmetry of 
the $(x_1,x_2,x_3)$ patch
 is explicitly broken by  $B_{tx_3}$ .) 
These  massless vacua are
constituted by D2-branes and F-strings only. This
 essentially indicates that a little or no effect of 
D8-branes (cosmological constant)  
is seen in the $z\gg z_0$ IR region! How did it actually
happen? It will be important  to understand how do the deformed 
 solutions \eqn{sol2a8} interpolate
between a massive type vacua (in UV) and the massless one (in IR). 
Let us understand the basic reasons behind this mechanism.
 
Actually the  $B_{\m\n}$ 
field in  the background eq.\eqn{sol2a8} plays a nontrivial
double role. A massive field 
carries additional degrees of freedom (modes) as compared 
to the ordinary field. We will show that
for the solutions \eqn{sol2a8} these degrees of freedom 
get separated out from each other in different regions of the spacetime.
 From eq.\eqn{sol2a8} we find
 that the string coupling
and  corresponding $B$ field  in  $z\sim \infty$ IR region become 
(leading behaviour)
\bea
&&e^\phi\sim g_a {z_0\over z}+\cdots
, ~~~~B_{ty}^{massless}\sim  
{ L^2 z_0^2\over q z^4}
+\cdots
\eea
and the corresponding expressions in $z\sim 0$ UV region are
\bea
&&e^\phi\sim g_a+\cdots
, ~~~~B_{ty}^{Higgsed}\sim  { L^2 \over q z^2}+\cdots
\eea
where dots indicate subleading terms.
These  expressions tell us that the massive (Higgsed) $B$ 
mode $\sim {1\over z^2}$ is dominant  in the UV Lifshitz vacuum, 
while the massless mode  $\sim{1\over z^4}$ 
 is relevant in the IR region. Their different scaling behaviors also
make them distinct and well segregated modes in two separate 
energy regimes. Therefore the $B$ field in  the
background \eqn{sol2a8} should only be thought of as  composite of these
two types of modes,
\be
B_{ty}=B_{ty}^{Higgsed}+B_{ty}^{massless}.
\ee
It is  meaningful 
because a massive tensor field carries more
degrees of freedom than the usual massless  tensor.
 The Higgsed $B$-mode decouples from the $z\gg z_0$ IR region 
whereas the massless $B$-mode takes over there, and
 exactly opposite phenomenon occurs towards  $z\ll z_0$ UV region.
Since it is the massless $B$ mode which explicitly depends on $z_0$,
it will carry the tag of an {\it excitation}. 
It is remarkable that it happens to be like this. 

Further, as an explicit check,
we shall demonstrate in the following that all $m$ dependent terms indeed 
decouple from the field equations, rendering the field dynamics 
being described by the massless fields  only, being dominant in the IR region.

\subsection{ Decoupling of mass and cosmological terms in the IR regime}

We demonstrate here that for the excited vacua \eqn{sol2a8} 
the mass terms as well as the cosmological
constant term in mIIA theory indeed decouple 
 from the IR dynamics. To determine this  
 the terms in  massive supergravity action are  evaluated. 
Using the IR expressions of fields  in \eqn{sol2a9} we get the following 
estimates
\bea 
&&(d\phi)^2\sim O({1\over \sqrt{z}}) \br
&& e^{-\phi} (d B)^2 \sim O({1\over \sqrt{z}}) \br
&& e^{\phi/2} (dC)^2 \sim O({1\over \sqrt{z}}) \br
&& m^2 e^{3\phi/2} (B_{\mu\nu})^2 \sim O({1\over z^{5\over2}}) \br
&& m^2 e^{5\phi/2}  \sim O({1\over z^{5\over2}}) 
\eea
We  clearly see that in these expressions valid for $z\to\infty$ region 
the last two $m$-dependent  terms  
indeed are vanishing as compared
to  first three  quantities. It is mainly
due to  weak string coupling, $e^\phi\sim {1\over z}\sim 0$, 
that the mass  and the cosmological terms
 cannot play a
significant role in the low energy dynamics of these fields.
Since these $m$-dependent  terms in 
the action (also in  equations of motion) become subleading and
irrelevant, the bulk dynamics  will be effectively  
 described by  ordinary type IIA  theory  in  deep IR. 
This  phenomenon 
 simply demonstrates that the decoupling of Higgsed mode of  $B_{ty}$
which although has got a strong pivot in the $z\sim 0$ region,
 but  the weak string coupling 
 renders it ineffective in the IR region.
Furthermore, {\it it is rather usual for QFT that massive modes get decoupled
 from a low energy  regime, but what is unusual here 
is that even the cosmological
constant term (which couples to dilaton only) 
gets decoupled from the  supergravity  in the deformed
 $Lif_4^{(2)}\times S^1\times S^5$ vacua!}
This demonstrates that massive type IIA theory is only 
an effective theory valid near
UV scale in the deformed
 $Lif_4^{(2)}\times S^1\times S^5$ solutions \eqn{sol2a8}. 
At  small scales, where the energy of the 
system is lower, the mass/cosmological
 terms  get screened out by a weak string coupling, 
and dynamically these terms become  subleading.
  
Let us now study the  UV regime close to $z\sim 0$.  The excited solution
\eqn{sol2a8} flows to  scale invariant Lifshitz background 
described by the fixed point solution \eqn{sol2a}. We
again evaluate the following expressions in   UV region  
\bea 
&&(\partial_\mu\phi)^2\sim 0 \br
&& e^{-\phi} (d B)^2 \sim O({1}) \br
&& e^{\phi/2} (dC)^2 \sim O({1}) \br
&& m^2 e^{3\phi/2} B^2 \sim O({1})\br 
&& m^2 e^{5\phi/2}  \sim O(1) .
\eea
Here we find that the mass  terms are of the same order 
as the kinetic terms of the fields. These  
covariants are also independent 
of holographic coordinate $z$, meaning that the theory is at  
the fixed point. Thus
the mass terms in Romans theory indeed play an important role 
in the field dynamics near  UV fixed point, where the vacuum is 
described by  $Lif_4^{(2)}\times S^1\times S^5$ geometry.
The typical scale where this crossover of the mIIA-oIIA theories 
(and the respective vacuas) 
happens is governed by $z_0$ scale.

\section{ $Lif_4^{(2)}\times S^1\times S^5$ with  
a flow towards UV}
Having discussed IR flows, we 
next present completely different solutions of massive type IIA theory which 
drive the flow  from  towards a 
relativistic UV vacua.\footnote{
 There however  already exist known   examples where the RG flow
 directly takes $Lif^{(3)}$
vacua (in the IR)  to  $AdS_5$ vacua in the UV, see \cite{hs1308}.}
The $a=2$ Lifshitz vacua allows following exact deformation 
\bea\label{sol2b10}
&&ds^2= L^2\left(- {dt^2\over  z^4f} +{dx_1^2+dx_2^2\over z^2}+{dz^2\over
z^2}  +{dy^2\over  q^2f} + d\Omega_5^2 \right) ,\br
&&e^\phi=g_a f^{-1/2},
 ~~~~~C_{(3)}= -{L^{3} \over g_a}
{1\over  z^4}dt\wedge dx_1\wedge dx_2, \br &&
B_{(2)}=  { L^2\over q  z^2}f^{-1}dt\wedge dy  \ ,
\eea 
where new function $$f= 1+{z_c^2\over z^2}, $$
the parameter $z_c$ corresponds to  switching on  irrelevant operators. 
These involve growing type of modes towards UV.  
The vacua \eqn{sol2b10} are  exact solutions of  mIIA theory. 
The  deformations  induce  running  of the string coupling 
which start flowing to  weakly coupled  ($g_a<1$) regime
 at the  UV end. Especially 
in  $z\sim 0$ region, where $f\approx {z_c^2\over z^2}$,
the  solution \eqn{sol2b10} becomes: 
\bea\label{sol2b11}
&&ds^2\simeq L^2\left(- {z_c^2dt^2\over  z^2} +{dx_1^2+dx_2^2\over z^2}+{dz^2\over
z^2}  +{z^2 dy^2\over  q^2 z_c^2} + d\Omega_5^2 \right) ,\br
&&e^\phi=g_a {z\over z_c},  ~~~~~C_{(3)}= -{ L^{3} \over g_a}
{1\over  z^4}dt\wedge dx_1\wedge dx_2, \br &&
B_{(2)}=  { L^2z_c^2\over q }dt\wedge dy  \ ,
\eea 
In these asymptotic (UV) solutions
the $z_c$ is arbitrary  and can be absorbed by redefining 
$g_a, t, y$. We remark that
 the UV vacua \eqn{sol2b11} are  well  described by  
ordinary type IIA theory. However, 
the $y$ direction in \eqn{sol2b11}, as being compact, will develop a   
 pinching type singularity
in the near UV region. But it remains stable because the interactions
are weakened due to weak coupling.  It would  be appropriate
 to study them by going over
to  type IIB T-dual background. It is rather straightforward to convince
 oneself that on the type IIB side the solution \eqn{sol2b11} 
has got a  nicer description
as deformed $AdS_5 \times S^5$. 

In the remaining section, we wish to make sure that
the mass terms in the theory completely decouple leaving behind a perfectly
massless vacua (governed by ordinary type IIA theory) at UV side. For this,   
using the asymptotic backgrounds \eqn{sol2b11}, in $z\sim 0$ region, 
we evaluate the following quantities
\bea 
&&(\partial_\mu\phi)^2\sim O({ \sqrt{z}}) \br
&& e^{-\phi} (d B)^2 \sim 0 \br
&& e^{\phi/2} (dC)^2 \sim O({ \sqrt{z}}) \br
&& m^2 e^{5\phi/2}  \sim O( z^{5/2}) \br
&& m^2 e^{3\phi/2} B^2 \sim O({ z^{5/2}}) 
\eea
We  clearly observe that the last two $m$-dependent  terms  
indeed become subleading as compared
to   the kinetic terms for $z\sim 0$. Once again
due to  weakened dilatonic coupling, i.e. $e^\phi\sim z\simeq 0$, 
the mass  and  cosmological constant  do not play 
any significant role in the UV dynamics of the fields.
Since the mass  terms in 
the action (and in equations of motion) become subleading and
irrelevant,  the bulk dynamics in the UV regime is  
better  described by  oIIA theory. 
 From eq.\eqn{sol2b10} one can find
 that the string coupling
and  corresponding $B$ field  in the $z\sim 0$ region become
\bea
&&e^\phi\sim g_a {z\over z_c}+\cdots
, ~~~~B_{ty}^{massless}\sim  
{ L^2 z_c^2\over q }
+\cdots
\eea
while  corresponding expressions in $z\sim \infty$ region are
\bea
&&e^\phi\sim g_a+\cdots
, ~~~~B_{ty}^{Higgsed}\sim  { L^2 \over q z^2}+\cdots
\eea
where dots indicate subleading terms.
These  expressions tell us that the  Higgsed $B$-mode $\sim {1\over z^2}$ 
is dominant  in the IR Lifshitz vacuum, 
while a constant (but massless) $B$-mode 
 is important in the boundary region. Their  power behaviors 
also make them distinct and well segregated modes in two separate 
energy regimes. Therefore the net $B$ field in  the
background \eqn{sol2b10} ought to  be thought of as  composite 
 mode: $B_{ty}=B_{ty}^{Higgsed}+B_{ty}^{massless}$.
 This  implies that there is explicit
 decoupling of  Higgsed  $B$-mode from UV dynamics, 
which although has a strong support in  the IR.

\section{Cascade of dynamical exponents along RG flow:  
$Lif_4^{(3)} \to Lif_4^{(2)}\to Lif_4^{(1)}$ }

The cascading Lifshitz theories are obtained in which dynamical exponent
jumps as a consequence of RG flow. These solutions are the 
combinations of previous two type of solutions. In the following 
we write down these vacua as solutions of mIIA theory.
  \bea\label{sol2c12}
&&ds^2= L^2\left(- {dt^2\over  z^4f} +{dx_1^2+dx_2^2\over z^2}+{dz^2\over
z^2}  +{dy^2\over  q^2f} + d\Omega_5^2 \right) ,\br
&&e^\phi=g_a f^{-1/2},
 ~~~~~C_{(3)}= -{L^{3} \over g_a}
{1\over  z^4}dt\wedge dx_1\wedge dx_2, \br &&
B_{(2)}=  { L^2\over q  z^2}f^{-1}dt\wedge dy  \ ,
\eea 
with function $f={z_c^2\over z^2}+1+ {z^2\over z_0^2}$, where $z_c, z_0$ 
are widely separated scales, $z_c \ll z_0$, but  otherwise
remain free parameters. An explicit compactification along $S^1$ (y-circle)
and $S^5$ produces the following 4-dimensional metric (Einstein) 
and a dilatonic scalar 
  \bea\label{sol2c12a}
&&ds^2_{Einstein}\sim L^2 f^{1\over 2}
\left(- {dt^2\over  z^4f} +{dx_1^2+dx_2^2\over z^2}+{dz^2\over
z^2} \right) ,\br
&&e^{2\phi_4}=g_4^2 f^{-1/2}, 
\eea 
and associated gauge field $A_{(1)}\sim  { L^2\over q  z^2 f}dt $. The 
four-dimensional coupling is given by
$g_4^2=g_a^2q/L$. Since the function $f\ge 1$ always,  the 
curvature of the spacetime still remains small
due to $L\gg 1$.
One can easily see that, in the UV region $z \ll z_c$,
 where $f\approx {z_c^2\over z^2}$, 
the solution \eqn{sol2c12a} reduces to the $a=1,~\theta=-1$ 
Lifshitz vacua 
  \bea\label{sol2c12b}
&&ds^2_{Einstein}\sim L^2 {z_c\over z}
\left(- {dt^2\over  z^2} +{dx_1^2+dx_2^2\over z^2}+{dz^2\over
z^2} \right) ,\br
&&e^{2\phi_4}\sim g_4^2 {z\over z_c}\ll g_4^2,~~~
 A_0\sim  { L^2\over q  z_c^2}
\eea 
discussed earlier.\footnote{
Here $\theta$ parameter stands for hyper-scaling (or effective conformal 
dimension) of the 4D Einstein metric as
per the convention:
 $ds^2_{Einstein}\sim  z^{\theta}
\left(- {dt^2\over  z^{2a}} +{dx_1^2+dx_2^2\over z^2}+{dz^2\over
z^2} \right)$}
Since the coupling gets weaker at higher energies, this phase of the
theory describe 
asymptotically free $3D$ Yang-Mills type (relativistic) theory on the boundary.
\footnote{ This CFT behaviour is similar to and is 
consistent with the UV behaviour of the known nonconformal
$CFT_3$ of the D2-branes.} While the ordinary D2-brane
theory in the IR flows to strongly coupled fixed point where M2-branes arise
\cite{itzhaki}. 
The difference in the present case arises in the IR regime. Due to 
nontrivial $A_0$, in
the IR regime the $CFT_3$ flows towards a Lifshitz  theory, 
as we will see next. 
At some intermediate scale $z=z_i$, such that 
$z_c\ll z_i\ll z_0$,  where $f\approx 1$, 
the solution \eqn{sol2c12a} would resemble with 
following $a=2,~\theta=0$ (scale invariant)
Lifshitz vacua,
  \bea\label{sol2c12c}
&&ds^2_{Einstein}\sim L^2 
\left(- {dt^2\over  z^4} +{dx_1^2+dx_2^2\over z^2}+{dz^2\over
z^2} \right) ,\br
&&e^{2\phi_4}\sim g_4^2, ~~~A_0\sim  { L^2\over q  z^2 } 
\eea 
It describes a critical point behaviour in a nonrelativistic Lifshitz
like theory. If we further lower the energies,
 in the deep IR region,
$z\gg z_0$,   where $f\approx {z^2\over z_0^2}$, 
 the solution \eqn{sol2c12a} 
becomes more like $a=3,~\theta=1$ Lifshitz vacua:
  \bea\label{sol2c12d}
&&ds^2_{Einstein}\sim L^2 {z\over z_0}
\left(- {z_0^2dt^2\over  z^6} +{dx_1^2+dx_2^2\over z^2}+{dz^2\over
z^2} \right) ,\br
&&e^{2\phi_4}\sim g_4^2 {z_0\over z}\ll g_4^2, 
~~~A_0\sim  { L^2z_0^2\over q  z^4 }. 
\eea
Thus in the $z\gg z_0$ IR region, the coupling gets weaker at 
farther  distances. Hence
this Lifshitz $a=3$ phase exhibits  `deconfinement' behaviour at  low energies, 
like usual electrodynamics. See the complete plot of the flow of the
coupling in figure \eqn{figu4b}. 
Due to  identical background fields  all these three Lifshitz behaviours can 
be characterized through:
 \bea\label{sol2c12e}
&&ds^2_{Einstein}\sim  ({z\over\lambda})^\theta
\left(- {dt^2\over  z^{2a}} +{dx_1^2+dx_2^2\over z^2}+{dz^2\over
z^2} \right) ,\br
&&e^{2\phi_4}\sim  ({\lambda \over z})^\theta, 
~~~A_t\sim  { \lambda^\theta\over   z^{a+\theta} }. 
\eea
where the gauge field is Maxwellian (for $a=3,~1$) and Proca type (for $a=2$).
The scale $\lambda$ defines the range of validity. For 
$z\ll \lambda$ the solutions describe $a=1,~\theta=-1$ UV relativistic region well,
whereas for
$z\gg \lambda$ the solutions describe $a=3,~\theta=1$ IR Lifshitz region. 
 \begin{figure}[ht]
\caption{ \label{figu4a} 
\it The figures are drawn for the cascading 
$Lif_4^{(a)}\times S^1\times S^5$. In the top figure 
the vertical size is indicative of physical radius of $y$-circle
in various $z$ regions.
The end stars indicate  possible $S^1$ pinching.
In the lower figure the same situation is 
viewed on  type-IIB theory side. 
The size of T-dual  $\tilde{y}$-circle is minimum at the throat,
 supported by constant $\chi$ flux. The dynamical effect of flux
 get diluted in the far IR and UV region. The geometry
appears like a traversable wormhole.}
 \centerline{ \epsfxsize=3in \epsffile{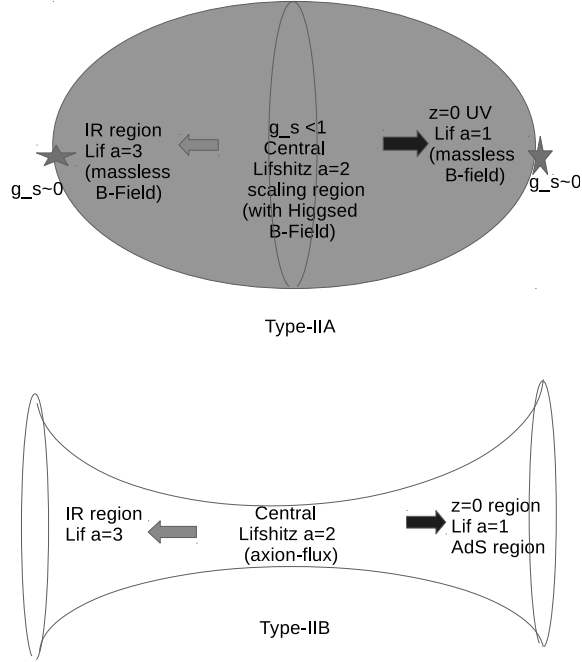} }
\end{figure}
Hence these composite kind of solutions can also be called the  Lifshitz
`wormholes' in which  $Lif_4^{(3)}$ 
world tunnels through a $Lif_4^{(2)}$ throat region
and finally emerging into a relativistic Yang-Mills phase as we  start
from  IR region and tune towards 
$z\sim 0$ region. An interesting aspect is that
the difference $$a-\theta=2$$ remains fixed along all three Lifshitz regions.
It is sort of a quantity which remains conserved all along the $z$-flow. 
\begin{figure}[ht]
\caption{ \label{figu4b} \it
 There is a fixed point at $z=z_i$ where coupling is maximum $g_4$.
But on the either side of it the coupling decreases. 
The two phases both of weak couplings, the `confining' Yang-Mills type 
(for $z_c\ne 0$)
and `deconfined' Lifshitz $a=3$ type (if $z_0\ne 0$)
are well separated.}
 \centerline{\epsfxsize=2.5in \epsffile{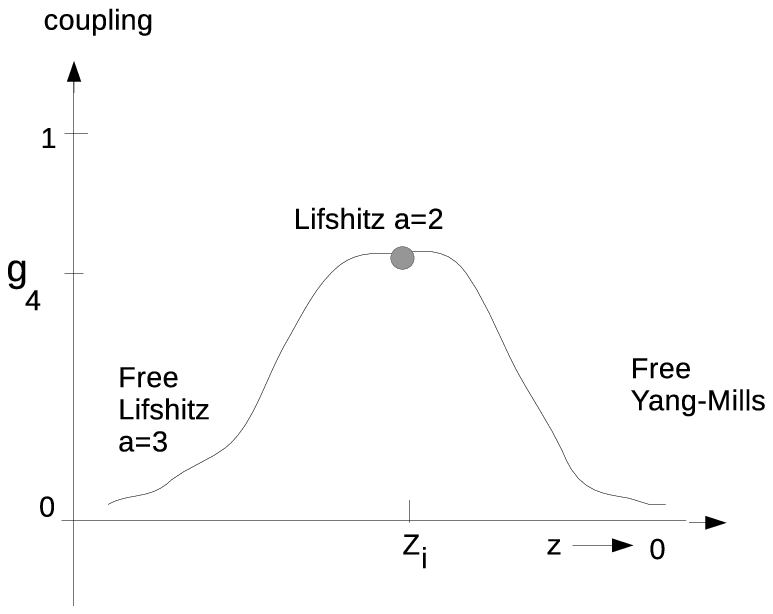} }
\end{figure}

To gain further insight let us evaluate the running of effective string coupling, 
$g_4\equiv e^{\phi_4}$,
with respect to scale $\lambda\equiv z$. This would indicate to us about the
nature of the flow of the coupling. We obtain it in the following way
\bea
&&\beta_{UV}={d g_4\over d\ln \lambda}\sim g_4, ~~~~{\rm for }~ z\sim 0 \br
&&\beta_{IR}={d g_4\over d\ln \lambda}\sim -g_4, ~~~~{\rm for}~ z\sim\infty 
\eea
We find that on the either side of the intermediate scale, 
$z_i$, the $\beta$'s change the sign. It implies
that $z_i$ would be a fixed point, where $\beta=0$. 
The fixed point corresponds to
$Lif_4^{(2)}$ spacetime of constant string coupling $g_4$. 
There is a resulting
scale invariance in the boundary  Lifshitz theory. 
The running string coupling is also sketched in the figure \eqn{figu4b}.

We comment that the physical size of $y$-circle 
becomes minimum near the two extremities; viz $z=\infty$ (horizon) and 
$z=0$ (boundary), where 
the associated string couplings become vanishing.
It means that the fluctuations  are stable and highly  suppressed 
and controlled. The configurations also preserve  some supersymmetry.
It can be better appreciated by going  
to a dual type-IIB set up 
near the two asymptotic regions. It is an straight forward analysis 
and we are avoiding full discussion here (one can see the   
solution given in the Appendix).
 Thus we indeed observe a cascade of Lifshitz theories with integral jumps in
the dynamical exponents.
 These basic ideas are depicted in the figure \eqn{figu4a}.
A few  observations can be
drawn from these cascading Lifshitz solutions 
involving  particular combination of background fields: 
\bi
\item The dynamical exponent $a$ decreases along the flow towards UV region. 
This behaviour is quite reasonable because covariance is expected to get
 restored at higher energies.
\item The associated hyperscaling parameter $\theta$ also has 
decreasing integral jumps towards higher energy scales.
 But the quantity $(a-\theta)$ is found to be conserved 
as we pass through various energy scales. 
\item At least  8 supersymmetries are intact in the IR asymptotic region. In 
the asymptotic UV region full supersymmetry of the relativistic
theory would be recovered.
\ei

\section{Summary}

We have presented
 new class of solutions for the  massive type IIA theory
under which massive string modes  become excited. The geometry is  
such that the Lifshitz $Lif_4^{(2)}\times {S}^1\times S^5$ vacua 
 emerges  near the UV fixed point of the RG flow.
In the deep  IR region these
solutions flow towards conformal
 $Lif_4^{(3)}\times S^1\times S^5$ background which belongs to
 ordinary type IIA theory. Thus the deep IR region is governed by
ordinary type-IIA theory where $B$ field is massless, 
while the UV region requires a massive $B$ field in massive type-IIA theory.
 We have explicitly verified that in the deep IR region 
all mass terms including the cosmological term
 indeed decouple from the field equations, rendering the whole
IR field dynamics in  $Lif_4^{(3)}\times S^1\times S^5$ vacua
controlled by the massless fields only. The latter vacua form 
 $1/4$-BPS states of ordinary type IIA theory. 
There is a crossover scale where the
swap in the theories  happens which is governed by the deformation scale
$z_0$. The crossover does happen for any $z_0$ value though. 
We have also presented 
another kind of solutions which describe a flow starting from 
$Lif_4^{(2)}\times {S}^1\times S^5$  IR vacua 
to $Lif_4^{(1)}\times {S}^1\times S^5$ relativistic vacua in  $z\sim 0$
 region. These relativistic vacua at the UV end point would be maximally 
supersymmetric.
In both of these RG flows the 
massive mode of the  $B$-field, particularly 
$B^{Higgsed}_{ty}\sim {1\over z^2}$, gets decoupled along with associated 
cosmological constant  term (of the same order) in the Roman's theory. 
We believe  primarily it happens due to vanishing string interactions 
with the massive mode, primarily the interactions being diminished
by vanishing  string couplings in  two asymptotic regions. 

We have also presented a full solution where  a
cascading flow of dynamical exponents 
$Lif_4^{(3)} \to Lif_4^{(2)}\to Lif_4^{(1)}$  is observed
 along the RG flow from IR to UV and vice versa.  
The $a=2,~\theta=0$ Lifshitz vacua in the center
flows to $a=3,~\theta=1$ and $a=1,~\theta=-1$  vacuas 
in the IR and UV regions respectively. Though
along the entire cascading flow the
quantity $(a-\theta)=2$ remains fixed. 
 The string coupling remains maximum  for the
$a=2$ solution, but it flows to weaker couplings on the either
side of it. The two phases of weak couplings, the `confining' Yang-Mills type 
in the UV (provided $z_c\ne 0$)
and a `deconfining' Lifshitz $a=3$ type in the IR (once ${1\over z_0}\ne 0$)
are well separated by a fixed point (critical) phase.  
Generically all physical (not ideal systems) situations would
 involve $z_c\ne 0,~{1\over z_0}\ne 0$ type fluctuations, no 
matter however small values these quantities take, 
the smooth flow described by our cascading
 solution appears to be realistic.  It would be interesting to explore
 $2d$ planar systems at ultra low energies  
where this picture might be realized.


\appendix{
\section{Cascading Lifshitz vacua in type-IIB theory}

For the completeness we note down the  
type-IIB string vacua with constant axion `flux'  which are 
`massive' T-dual  of the  Lifshitz 
solutions \eqn{sol2c12} in massive type IIA theory.
The background is
\bea\label{sol2c12n}
&&ds^2= L^2\left(- {2 dt d {y}\over z^2} +{g\over z^2}
  d {y}^2  +{dx_1^2+dx_2^2\over z^2}+{dz^2\over
z^2}  + d\Omega_5^2 \right) ,\br
&&e^\phi=g_b, ~~~\chi={2 \tilde{q}\over g_b}  y\ , 
\eea 
supported by  self-dual 5-form $F_{(5)}$ field strength. 
 The function $g$ is given by  $g(z)={\tilde{z}_c^2}+\tilde{q}^2 z^2+ 
{z^4\over \tilde{z}_0^2}$. 
The new constants  $\tilde{z}_c, \tilde{z}_0$ 
are two far separated scales, $\tilde{z}_c \ll \tilde{z}_0$, but they 
remain free parameters in type IIB also. But the parameter
$\tilde{q}$ is tightly related to the axion field. 
The type IIB/A string couplings
are related as $g_b={g_a q\over L},~\tilde{q}=m g_b/2$ etc. 
A  Kaluza-Klein 
reduction along $S^1\times S^5$ 
will produce  four-dimensional solutions 
same as given in \eqn{sol2c12a}. 
 It is well known that \cite{berg}, because
 the axionic  field strength 
$F_{(1)}=d\chi=constant$, these constant fluxes 
produce massive supergravity theories under generalized
Scherk-Schwarz compactification on $S^1$ \cite{scherkschwarz,berg,llp}. 
\subsection{Supersymmetries in the cascade}
The supersymmetry of these type IIB background 
can be found by evaluating the fermionic variations in type IIB theory. 
These variations can be obtained from
 the works \cite{berg, schwarzwest}. To simplify the effort,
 we will set $L=1, g_b=1$. From eq.\eqn{sol2c12n}
the vielbeins are defined as
\bea
e^+={1\over z} (dt -  g dy), 
~~e^-= {1\over z} dy,   
~~e^1= {1\over z} dx^1,   
~~e^2= {1\over z} dx^2,   
~~e^3= {1\over z} dz, \cdots   
\eea
so that ten-dimensional line element becomes 
$$ds^2=-2 e^+ e^-+e^1e^1+e^2e^2+e^3e^3+\cdots$$ 
and the self-dual 5-form as: 
$F_{(5)}\sim (1+ \star_{10}) e^+\wedge e^-\wedge e^1\wedge e^2\wedge e^3$.
The dots imply  similar expression for $S^5$. 
Using the differential geometric identity 
$de^a + \omega^a_b \wedge e^b=0$, the spin connection 
1-forms can be evaluated. These mostly will be of the same type as in 
the case of exact $AdS_5\times S^5$ geometry, 
except the following connection component
\bea
\omega^+_{~3}=-{1\over z} dt - z\partial_z ( {g\over z}) dy  
\eea
which has a new contribution from $g$ dependent term. 
Now when we evaluate the dilatino variation
for the above background, it reduces to
\bea
0=\delta \lambda= \partial_y \chi \Gamma^y \epsilon=2 \tilde{q} e^y_-\gamma^-\epsilon
\eea
The vanishing fermionic variations put a rigid 
condition on the Killing spinors that 
\be \label{gh8}
\gamma_+\epsilon=0
\ee
At this stage we have chosen 
the Killing spinors to be precisely that of  anti de Sitter spacetime
$\epsilon\equiv \epsilon_{AdS\times S}$. 
(Note $(\gamma_+)^2=0$, and $\gamma_a$'s are
undressed  gamma matrices.) Such a restriction
would break all sixteen supernumerary Killing spinors of AdS spacetime. 
The condition however will  allow eight Poincare (ordinary) type Killing
spinors remaining intact. 

We next have to evaluate the gravitino variations:
$\delta \Psi_\mu=0$, to obtain further conditions, if any. We keep 
in mind that $\omega_{,y}^{+3}$ is nontrivial and  given the
condition \eqn{gh8}.  
We do not present full details of these calculations here, 
but all gravitino equations will get satisfied
except the following one 
\bea
0=\delta\Psi_y= (\partial_y+{1\over 4}\omega^{ab}_{,y}\gamma_{ab})
\epsilon +{i\over 4} \epsilon \partial_y\chi -{i\over 192} \Gamma^{(5)} 
\Gamma_y F_{(5)}\epsilon
\eea 
which needs to be evaluated separately as it contains 
$g$ dependent contribution.
Substituting the background fields from \eqn{sol2c12n}, 
this simplifies to 
\bea
\partial_y\epsilon+{i \tilde{q}\over 2} \epsilon=0 .
\eea
That has an immediate solution of the type: $\epsilon =e^{-i\tilde{q}y\over 2 } 
\epsilon_{AdS\times S}$, along
with the condition \eqn{gh8}. Thus the Killing
 spinors will also have explicit $y$ dependence, but
no new condition is required. When $\tilde{q}=0$ this dependence on $y$
will drop out. However, the condition $\gamma_+\epsilon=0$ will still
be there so long as  $z_0'$ is nontrivial.
Thus the  supersymmetry count for the background \eqn{sol2c12n} reduces 
to 8 Killing spinors being intact. We expect at least these many
supersymmetries will survice if we T-dualise \eqn{sol2c12n} 
back to get mIIA cascading solution \eqn{sol2c12}. The explicit 
$y$ dependence in the Killing spinors is induced only due to the axion flux. 
Generically a massive T-duality preserves supersymmetry. 
It is not clear whether these Killing spinors would survive after
compactification. However, if the axion flux is switched off, then there
are certainly 8 Killing spinors. Thus we comment that, in the 
two asymptotic regions where the axion flux is sufficiently weakened the 
supersymmetry will be regained. In conclusion, we have got 
a supergravity vacua where supersymmetry can be gained or  lost
dynamically, both in the IR region as well as in the UV regime.

\end{document}